\documentclass[aps,prl,preprint,superscriptaddress,showpacs,preprintnumbers]{revtex4}

\usepackage{graphicx}
\usepackage{dcolumn}

\begin{document}
\title{A Composite Chiral Pair of Rotational Bands in the odd-A Nucleus $^{135}$Nd}
\author{S. Zhu}
\author{U. Garg}
\author{B.K. Nayak}
\affiliation{Physics Department, University of Notre Dame, Notre Dame, IN 46556}
\author{S.S. Ghugre}
\author{N.S. Pattabiraman}
\affiliation{IUCDAEF-Calcutta Center, Calcutta 700 094, India}
\author{D.B. Fossan}
\author{T. Koike}
\author{K. Starosta}
\author{C. Vaman}
\affiliation{Department of Physics and Astronomy, State University of New York, Stony Brook, NY 11794}
\author{R.V.F. Janssens}
\affiliation{Physics Division, Argonne National Laboratory, Argonne, IL 60439}
\author{R.S. Chakrawarthy}
\author{M. Whitehead}
\affiliation{Schuster Laboratory, University of Manchester, Manchester M13 9PL, UK}
\author{A.O. Macchiavelli}
\affiliation{Nuclear Physics Division, Lawrence Berkeley National Laboratory, Berkeley, CA 94720}
\author{S. Frauendorf}
\affiliation{Physics Department, University of Notre Dame, Notre Dame, IN 46556}
\date{\today}

\begin{abstract}
High-spin states in $^{135}$Nd were populated with the
$^{110}$Pd($^{30}$Si,5n)$^{135}$Nd reaction at a $^{30}$Si
bombarding energy of 133 MeV. Two $\Delta$I=1 bands
with close excitation energies and the same parity were oberved. These
bands are directly linked by $\Delta$I=1 and $\Delta$I=2 transitions.
The chiral nature of these 
two bands is confirmed by comparison with three-dimensional tilted axis
cranking 
calculations. This is the first observation of a three-quasiparticle chiral
structure and establishes the primarily geometric nature of this phenomenon.
\end{abstract}


\maketitle

\begin{figure}
\includegraphics[trim=1.8cm 12.6cm 6cm 10cm,scale=0.36,angle=-90]{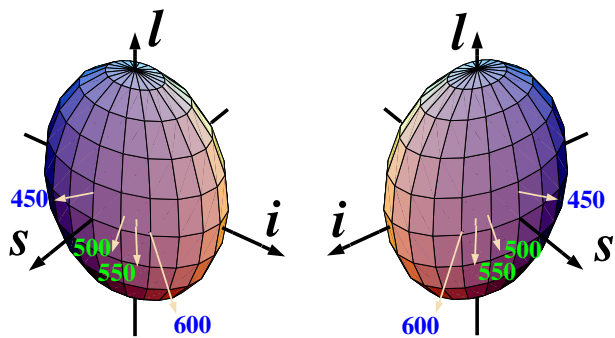}
\vspace*{-1.5cm}
\caption{\label{fig:orient} The orientation of the total angular
momentum vector, $I$, with respect to the axes of the triaxially 
deformed density distribution in $^{135}$Nd (light arrows) for different
rotational
frequencies from the tilted-axis cranking calculations. The numbers show
the rotational frequency in keV. The labels l, s and i stand for long-, short-
and intermediate-axis, respectively. In the order s-i-l, these axes form a
``right-handed'' system (left part of the figure) and a ``left-handed'' system
(right part of the figure)---a chiral doublet.}
\end{figure}

The rotational motion of triaxial nuclei  attains a chiral character if 
the angular momentum has substantial projections on all three principal axes 
of the triaxial density distribution \cite{rmp}. Fig.~\ref{fig:orient} 
illustrates how chirality
emerges from the combination of the triaxial geometry with an axis of 
rotation that lies out of the three symmetry planes of the ellipsoid. 
The three components of the angular momentum vector form either 
a left-handed or a right-handed system. Reverting the
direction of the component of the angular momentum on the intermediate
axis changes the chirality. The left-handed and right-handed configurations,
which have the same energy, manifest themselves as two degenerate rotational
bands---the chiral doublet. This argument, which is based only on the symmetry
of the rotating triaxial nucleus, is independent of how the three components 
of the angular momentum are composed \cite{rmp}. The simplest possibility was
considered in Ref.~\cite{frau1} where the concept of chiral 
rotation was first 
suggested: one proton aligns its angular momentum with the short axis, one
neutron hole aligns its angular momentum with the long axis, and the collective
angular momentum generated by all other nucleons aligns with the 
intermediate axis. 
Experimental evidence for this simplest type of a chiral configuration was
first found in $^{134}$Pr \cite{petrache} and subsequently in a number of 
other odd-odd nuclei \cite{starosta3,hecht,koik,hartley,starosta1,beausang,bark,fossan}. 
However, chirality is expected for {\em all} configurations that have
substantial angular momentum components along the three  principal axes,
no matter how the individual components are composed. In this  Letter,
we report the first observation of a pair of chiral bands in an odd-A
nucleus $^{135}$Nd. In this case, the angular
momentum of {\em two} protons is aligned with the short axis and, thus, the
chiral bands are based on a three-quasiparticle configuration. Our
results represent an important confirmation of the geometrical interpretation
in terms of broken chiral symmetry \cite{rmp}, which claims that pairs of nearly
degenerate $\Delta$I=1 bands with the same parity appear whenever there is a 
chiral geometry of the angular momentum components, irrespective of how
they are composed. 

High spin states in $^{135}$Nd were populated via the
$^{110}$Pd($^{30}$Si,5n)$^{135}$Nd reaction, at a bombarding
energy of 133 MeV. The incident beam energy was optimized for this
reaction channel by measuring an excitation function. For the
purpose, $^{30}$Si beams at four
different energies from 128 MeV to 143 MeV were obtained in steps of 5 MeV 
from the FN-tandem injected superconducting LINAC at the State University
of New York at Stony Brook.
Three Compton-suppressed HPGe detectors were used for $\gamma$ 
detection. The beam for the coincidence measurements was provided by the 88-inch
cyclotron facility at the Ernest O. Lawrence Berkeley National
Laboratory. Two stacked, self-supporting, isotopically enriched
target foils ($\sim$0.5 mg/cm$^{2}$ thick) were used. Quadruple-
and higher-fold coincidence events were measured using the
Gammasphere array \cite{gamma} in stand-alone mode, which comprised 103 active
Compton-suppressed HPGe detectors. A total of about
1.3$\times$10$^{9}$ events were accumulated and stored onto
magnetic tapes for further analysis.
The data were sorted into three-dimensional (E$_\gamma$-E$_\gamma$-E$_\gamma$ 
cube) and four-dimensional (E$_\gamma$-E$_\gamma$-E$_\gamma$-E$_\gamma$ 
hypercube) histograms using the RADWARE formats \cite{rad1}. These were 
analyzed with the Radware software package, which
uses the generalized background subtraction algorithm of
Ref.~\cite{rad2}, to extract ``double-gated'' and ``triple-gated'' spectra. 
The level scheme was constructed by examination of these gated spectra. The
multipolarity assignments were made on the basis of DCO ratios \cite{kra}
and confirmed by an angular distribution analysis \cite{iacob}.

\begin{figure}[t]
\includegraphics[trim=2cm 5.5cm 2cm 3cm,scale=0.53]{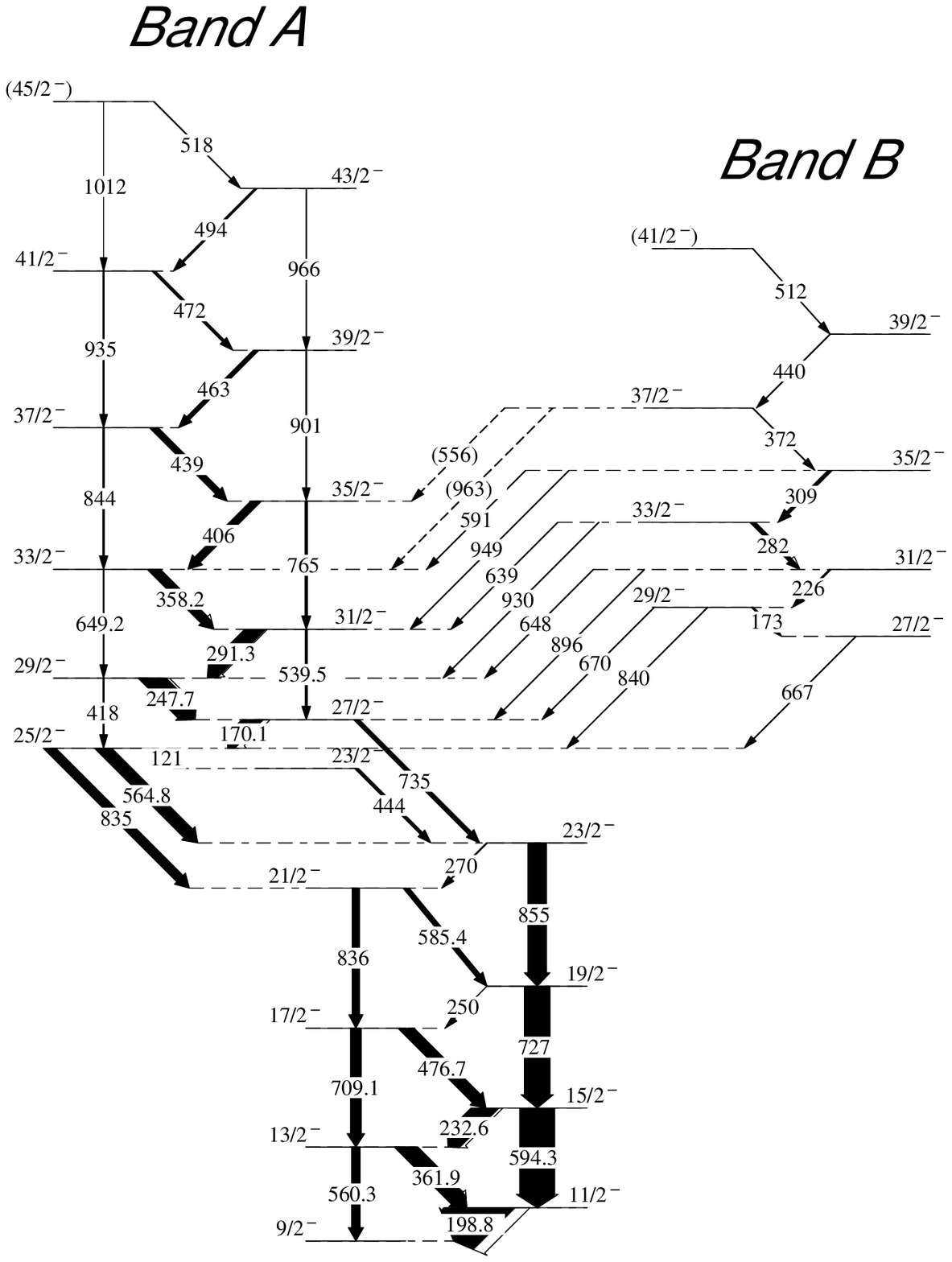}
\caption{\label{fig:level} Partial level scheme showing the
$\pi$h$_{11/2}^{2}\nu$h$_{11/2}^{-1}$ band and the newly observed
sideband of $^{135}$Nd. Bands A and B are interpreted as the chiral pair. The transition intensities are proportional to the thickness of the arrows.}
\end{figure}

\begin{figure}[t]
\includegraphics[trim=1.5cm 8.5cm 0cm 4cm,scale=0.45]{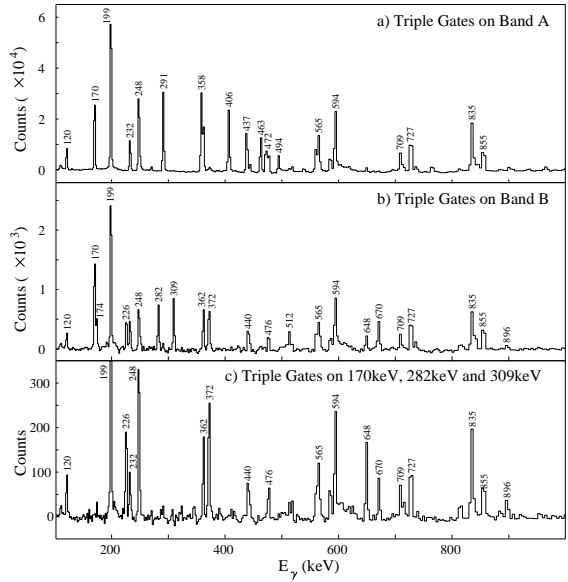}
\vspace*{1.5cm}
\caption{\label{fig:spectra} Background subtracted
$\gamma-\gamma-\gamma-\gamma$ spectra in $^{135}$Nd from the
summation of spectra with triple gates set on band A (top panel) and B
(middle panel). The bottom panel is included to clearly show the presence of
the 648-keV ($\Delta$I=1), 670-keV ($\Delta$I=1), and 896-keV ($\Delta$I=2)
``linking'' transitions in the gated spectra.}
\end{figure}

A partial level scheme for $^{135}$Nd, as obtained from the present work, is
shown in Fig.~\ref{fig:level}. The primary feature of interest in the level
scheme is a new $\Delta$I=1 band (band B) which feeds into, and
forms a chiral pair with, the structure identified as band A in the
yrast sequence. The strong links between the two bands point to their same
intrinsic configuration.
Fig.~\ref{fig:spectra} shows the summed gated coincidence spectra for bands A
and B. The spins and parity of the yrast band have been assigned in previous
work \cite{piel,beck}; these agree with the assignments extracted
from the aforementioned DCO and angular distribution analyses. The spins in 
band B have been extracted in exactly the
same manner. The $\Delta$I=1 character of the intraband transitions in band B 
and of the transitions linking it with band A is also confirmed by 
angular distribution measurements---typical values of the $\frac {A_2}{A_0}$
and $\frac {A_4}{A_0}$ 
coefficients for these transitions are $\sim -$0.4 and $\sim$0, respectively.
In particular, the $\frac {A_2}{A_0}$ratios for the 648- and 670-keV linking
transitions are -0.654(30) and -0.564(22), respectively, establishing their
$\Delta$I=1 character. 
To ascertain the mixed M1/E2 multipolarity 
of the intraband transitions, electron conversion (EC) coefficients were
estimated from intensity-balance arguments, {\em i.e.}, by 
comparing the intensities of the $\gamma$ rays feeding into and decaying out 
of the levels involved. For Z=60
and E$_\gamma$=250 keV, the EC coefficient is $\sim$ 0.2 for an M1 transition
and $\sim$ 0.02 for an E1 transition. The values of the EC coefficients
extracted 
from the data for the 170- and 226-keV transitions are 0.25(10) and 0.17(3),
respectively, clearly pointing to their M1/E2 character. 
Looking at the linking transitions, the calculated DCO ratio for an
E1 transition in our geometry is $\sim$ 0.9, while the measured 
values for the DCO ratio for the 649- and 670-keV, $\Delta$I=1 linking
transitions are $\sim$ 1.5, pointing to their M1/E2 character as well. While
it has not been possible to extract angular distribution ratios for the other
linking transitions because of their weak intensities, the ratio of intensity
at forward angles to that at 90$^{\circ}$ for the 896-keV transition is 1.7(3),
establishing it as an E2 transition. It is evident, therefore, that the two
bands have the same parity and are linked by mixed M1/E2 and E2 transitions.

\begin{table}
\caption{\label{tab:table1}Orientation angles $\theta$, $\phi$, values
$I(\omega)$ of
the angular momentum, and the excitation energies for the configuration
[$\pi h_{11/2}^{2}, \nu h_{11/2}^{-1}$] in $_{\ 60}^{135}$Nd from the 3D TAC
calculations described in the text. Chiral solutions correspond to both
$\theta \neq 0^{\circ}$ and 90$^{\circ}$ and
$\phi \neq 0^{\circ}$ and 90$^{\circ}$. The angles $\theta$ and $\phi$ are
defined with respect to the $l$- and $s$- axes, respectively (see Fig. 1).}
\begin{ruledtabular}
\begin{tabular}{cccc}
$\hbar\omega$[MeV]&$\theta$&$\phi$&$I(\omega)$[$\hbar$]\\
\hline
0.20&60&0&16.5\\
0.25&65&0&17.4\\
0.30&65&0&18.1\\
0.35&65&0&18.8\\
0.40&65&0&19.5\\
0.45&65&0&20.1\\
0.50&75&$\pm$28&22.0\\
0.55&75&$\pm$37&25.0\\
0.60&85&$\pm$47&30.4\\
\end{tabular}
\end{ruledtabular}
\end{table}

Three-dimensional tilted axis cranking (3D TAC) calculations have been performed 
to investigate the possibility of chiral solutions in $^{135}$Nd. 
The TAC model has been discussed in detail in Refs.~\cite{frau3,frau2,frau4,dimit1,dimit2}. The deformation
parameters $\epsilon$ and $\gamma$, and the tilt angles $\theta$
and $\phi$ are obtained for a given frequency $\hbar\omega$ and for a given
configuration, by searching for a local minimum on the multi-parameter
surface of the total Routhian. This procedure led to a solution 
for the $\pi$h$_{11/2}^2\nu$h$_{11/2}^{-1}$ configuration with
$\epsilon$ and $\gamma$
values of $\sim$ 0.20 and $\sim$ 30$^\circ$, respectively, for frequencies 
$\hbar\omega$ between 0.20 MeV and 0.60 MeV. We include the neutron pairing correlations
($\Delta$=1.08MeV) but use zero pairing for the proton system, because the pair correlations
are blocked by the {\it two} h$_{11/2}$ protons. Table~\ref{tab:table1} gives 
detailed results of the calculation and Fig.~\ref{fig:orient} illustrates 
how the angular momentum vector
reorients with respect to the deformed density distribution. 

\begin{figure}[t]
\includegraphics[trim=8cm 5.5cm 6cm 6cm,scale=0.45]{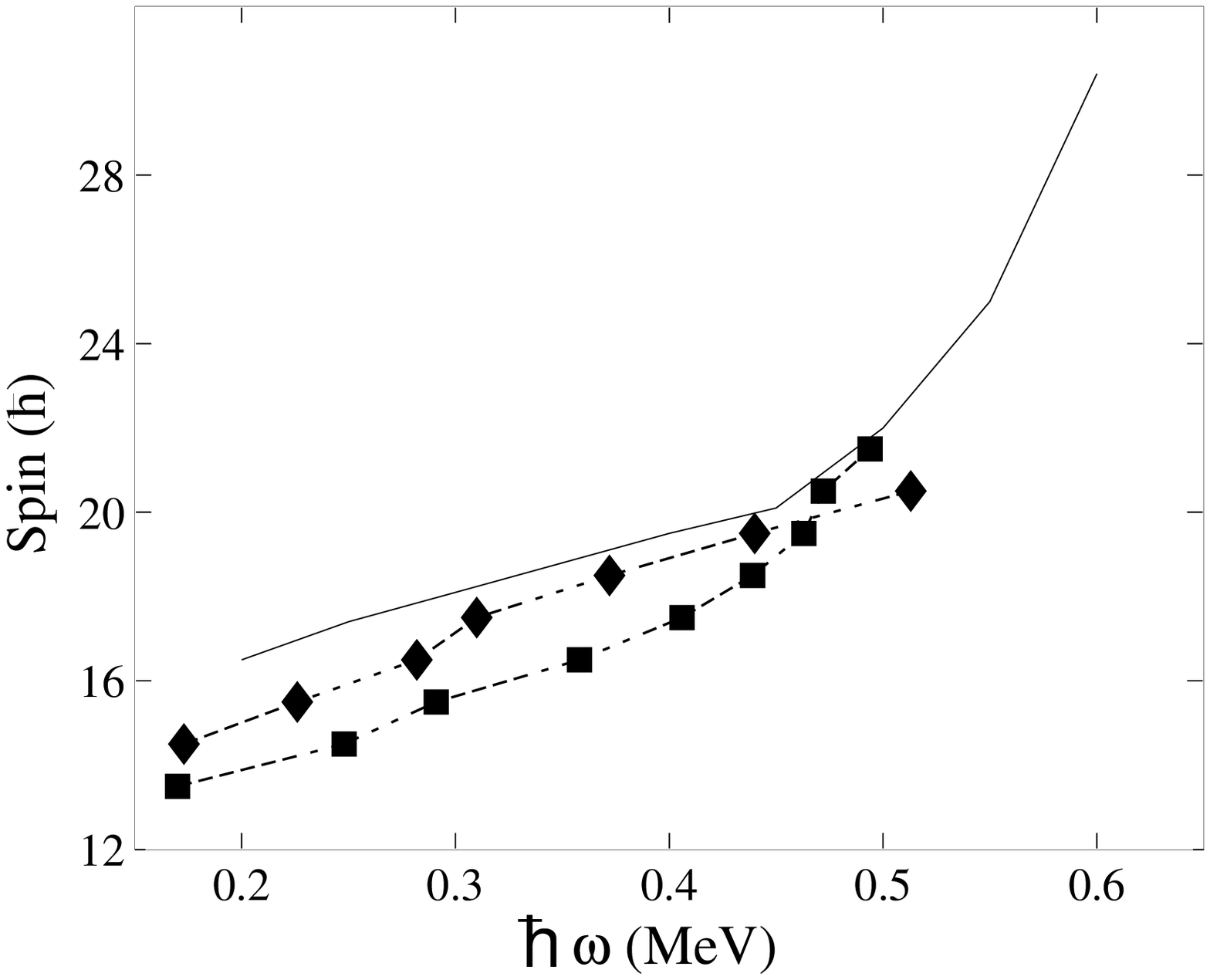}
\caption{\label{fig:omega} The experimental spins versus $\hbar\omega$ plots
for band A (squares) and band B (diamonds). The solid line is
the result of 3D TAC calculations.}
\end{figure}
\begin{figure}[t]
\includegraphics[trim=5.5cm 6.5cm 6cm 6cm,scale=0.45]{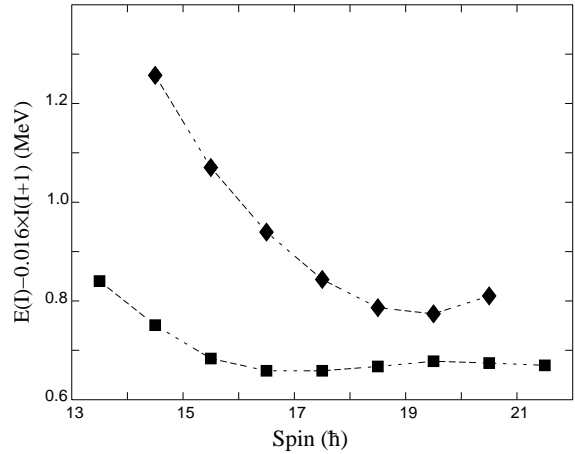}
\caption{\label{fig:energy} Experimental excitation energies versus spin for
band A (squares) and band B (diamonds). A rigid rotor reference has been
subtracted from the energies.}
\end{figure}

The ground band based on $\nu$h$_{11/2}$ has been studied
previously \cite{piel,beck}. It was suggested that the backbending
at I$^\pi$=$\frac{25}{2}^{-}$ is due to
the alignment of the lowest pair of h$_{11/2}$ protons, thereby giving  
the configuration $\pi h_{11/2}^2 \nu h_{11/2}^{-1}$.
Above the backbend, the yrast band and the new side band behave as expected
from the calculations. At first, the TAC solution is planar ($\phi=0$). At
$\hbar\omega\approx 0.45 MeV, I(\omega)\approx 20$, it becomes chiral.
The two bands are separated by about 400 keV for $I=\frac{29}{2}$. 
They approach each other
with increasing spin and the energy separation becomes as small as 100 keV
at $\hbar\omega\approx .45 MeV, I\approx 19$, which we interpret as the onset
of chirality. This frequency agrees well with the TAC calculation. 
The change from the planar to the chiral solution
is accompanied by a change of the slope of $I(\omega)$. This indicates
that, in the chiral regime, the angular momentum increases along the
intermediate axis whereas, in the planar regime,
it remains in the plane spanned by the short and long axes which have  
smaller moments of inertia.
In our data, band A shows such a change of slope at the expected frequency.
However, the corresponding change is not seen in band B (see below for an
explanation).  

For further interpretation, one has to take into account that 
TAC gives only the classical orientation, around which $\vec I$
executes 
a quantal motion. In the planar regime ($\phi=0$), these are slow oscillations 
of $\vec I$ into the left- and right-handed sectors, which are symmetric 
to the l-s plane; they have been
called chiral vibrations \cite{starosta3}. Band A, then, is the zero- and
band B the one-phonon state. The frequency of these vibrations decreases with 
angular momentum until the motion becomes unstable against static chirality. 
This decrease is reflected as a difference of about 1.5 units of 
angular momentum in Fig.~\ref{fig:omega}. In the chiral regime, there  
still remains some tunneling between the left- and right-handed configurations, 
which causes the remaining energy difference of about 100 keV between 
the bands. 

The energy difference between the $\frac {41}{2}$ states is slightly larger
than for 
the adjacent lower-spin states as shown in Fig.~\ref{fig:energy}. Such an 
increase of the splitting is expected from our TAC calculations. As seen 
in Fig.~\ref{fig:orient} and Table~\ref{tab:table1}, $\vec I(\omega)$
approaches 
the s-i plane for $\hbar\omega > 0.55 MeV$, which means that the tunneling between 
the left- and right-handed sectors through the s-i plane increases and so does
the splitting. For $\hbar\omega = 0.6 MeV$, the TAC solution is almost planar
and one is back in the regime of chiral vibrations. Static chirality is, thus,
a transient phenomenon, which appears only in a short interval
around $I=19$. The increase of the splitting at high spin may be the reason 
why the  expected upbend of the function $I(\omega)$ is not seen in 
the upper band---the highest-spin levels in band B are progressivly pushed 
higher in energy compared to those in band A, shifting the data points for
these levels in band B to the right in Fig.~\ref{fig:omega}.

Interestingly, a candidate for chiral-doublet bands has recently been proposed
in the even-even nucleus $^{136}$Nd as well \cite{pet2}; the suggested
configuration was
[$\pi(h_{11/2}, d_{5/2}/g_{7/2}), \nu(h_{11/2}^2)$]. However, in contrast
with the case reported here, and with all odd-odd neighboring nuclei where
chiral bands have so far been observed, the bands reported in $^{136}$Nd are
closest to each other in energy at the lowest spins, with the energy-separation 
increasing monotonically with increasing spin.
It is possible that the reported band-pair originates from a proton pseudo-spin 
doublet in the configuration [$\pi(g_{7/2}/g_{9/2},h_{11/2}), \nu(h_{11/2}^2)$]. 
Indeed, the energy differences are similar to a pair of
bands reported recently in $^{128}$Pr \cite{pet3}, where such an interpretation 
has been put forward.

In summary, chiral twin bands based on three quasiparticles have been observed
for the first time in this experiment. 3D TAC calculations reproduce the
experimental results well, confirming the underlying structure of these bands.
As a general property, when angular momentum components align with the three
principal axes in a $\gamma$-deformed system,
chirality manifests itself not only in the odd-odd systems, but also in 
odd-A systems. The present results establish the primarily geometric character
of this phenomenon.

This work has been supported in part by the U.S. National Science Foundation
(grants number PHY-0140324 and INT-0115336), the Department of Science and
Technology, Government of India, the U.K. Science and Engineering Research
Council, and the U.S. Department of Energy Nuclear Physics Division, under
contract no. W-31-109-ENG-38.

\bibliography{nd135prl}
\end{document}